\documentclass[aps,prb]{revtex4}
\usepackage{graphicx,amsbsy,bm}

\begin{document}

\title{Undamped energy transport by collective
surface plasmon oscillations along metallic nanosphere chain}
 \author{W. Jacak$^1$, J. Krasnyj$^{1,2}$, J. Jacak$^1$, A. Chepok$^2$, L. Jacak$^1$, W. Donderowicz$^1$, D. Z. Hu$^3$, and D. M. Schaadt$^3$}

\affiliation{$^1$Institute of Physics, Wroc{\l}aw University of Technology, Wyb. Wyspia\'{n}skiego 27, 50-370, Wroc{\l}aw, Poland;\\
$^2$ Theor. Phys. Group., International  University, Fontanskaya Doroga 33, Odessa, Ukraine;\\
$^3$ Institute of Applied Physics/DFG-Center for Functional Nanostructures, University of Karlsruhe, Karlsruhe, Germany}

\begin{abstract}

 The random-phase-approximation semiclassical scheme for description of plasmon excitations in large metallic nanospheres (with radius 10--100 nm)  is
 developed for a case of presence of dynamical electric field.
 The spectrum of plasmons in metallic nanosphere is determined including both  surface and volume type
 excitations and  their mutual connections. It is demonstrated that only surface plasmons of dipole type can be excited
 by a homogeneous dynamical electric field. The Lorentz friction due to irradiation of  e-m energy by
 plasmon oscillations  is analysed with respect to the sphere dimension. The resulting shift of
 resonance frequency  due to plasmon damping
 is compared with experimental data for various sphere radii.  Collective of wave-type oscillations of
 surface plasmons in long chains of metallic spheres are described. The undamped region of propagation of
 plasmon waves along the chain is found in agreement with some previous numerical simulations.

\end{abstract}
\vspace{1mm}
\vspace{1mm}

\maketitle

\section{Introduction}

Experimental and theoretical invesitigations of plasmon excitations in metallic nanocrystals rapidly grew
mainly due to perspectives of possible applications in photovoltaics and microelectronics.
A significant enhancement of absorption of the incident light in photodiode-systems with active surface
covered with nano-dimension metallic particles (of Au, Ag or Cu) with planar density $\sim 10^8$/cm$^2$ was observed\cite{wzmocn1,wzmocn2,wzr1,wzr2,wzr3,wzr4,wzr5}.
 This is
due to a mediating role in light energy transport played by surface plasmon oscillations in metallic nano-compounds on semiconductor surface.
These findings are of practical importance towards enhancement of solar cell efficiency especially for thin film cell technology. Hybridized states
of the surface plasmons and photons result in plasmon-polaritons\cite{maradudin} which are of high importance for applications
in photonics and microelectronics\cite{zastos,plasmons}, in particular  for transportation of energy in metallic modified structures in nano-scale\cite{bb,dd}.

Surface plasmons in nanoparticles were widely investigated since their classical description by Mie\cite{mie}.
Many particular studies, including numerical modelling of multi-electron clusters, have been carried out\cite{brack1,serra,brack,ekart,ksi}. They were
developments of Kohn-Sham attitude in form of LDA (Local Density Approach) or TDLDA (Time Dependent LDA)\cite{brack1,brak,serra,ekart}
addressed, however, to small metallic clusters, up to ca. 200 electrons (as limetted by severe numerical constraints).
 The random phase approximation (RPA) was formulated\cite{rpa} for description of volume plasmons in bulk metals
 and utilised also for confined geometry mainly in numerical  or semi-numerical manner\cite{brack}. Usually, in these analyses the {\it jellium}
 model was assumed for description of positive ion background in metal and dynamics was addressed  to electron system only\cite{brack,ekart,ksi},
 and such an attitude is preferable for clusters of simple metals, including noble metals (also transition and alkali ones).

In the present paper we generalise the bulk RPA description\cite{rpa}, using semiclassical approach, for a large metallic nanosphere (with radius of several tens nm, and with $10^5--10^7$
electrons)
 in an all analytical calculus version\cite{jac}. The plasmon oscillations of compresional and traslational
 type, resulting in excitations  inside the sphere and on its surface, respectively, are analysed
 and referred to volume and surface plasmons.
 Damping effects of plasmons via electron scattering processes and radiation losses are included, the latter ones, via Lorentz friction force.
 The shift of the resonance frequency of dipole-type surface plasmons (only plasmons  induced by homogeneous time-dependent
 electric field), due to damping phenomena, is compared with the experimental data for various
 nanosphere radii.
  Collective surface dipole-type plasmon oscillations in the linear chain of metallic nanospheres are analysed
 and wave-type plasmon modes are described. A coupling in near field regime between oscillating dipoles of surface plasmons
 together with retardation effects of energy irradiation lead to a possibility of undamped propagation
 of plasmon waves along the chain in the experimentally realistic region of parameters (of separation of spheres in the chain and their radii).
 These effect would be of particular significance for by plasmon arranged transport of energy along metallic chains for
 application in nanoelectronics.

 The paper is organised as follows. In the next section the standard RPA theory in quasiclassical limit,
 is generalised for the confined system of spherical shape.
 The resulting equations for volume and surface plasmons are solved in the following section
 (with particularities of calculus shifted to the Appendix).
 The next section contains  description of the Lorentz friction for surface plasmons oscillations of the dipole-type. The analysis of
 the collective wave-type surface plasmon oscillations in the chain of metallic nanospheres is presented in the last section.
 Besides the theoretical model the comparison of the characteristic nano-scale plasmon behaviour with available experimental data,
 including own measurements, is presented.

\section{RPA  approach  to electron excitations in metallic nanosphere}
\subsection{Derivation of RPA equation for local electron density in a confined spherical geometry}

Let us consider a metallic sphere with a radius $a$
located in the vacuum, $\varepsilon = 1,\;\mu =1$ and in the presence of dynamical electric field (magnetic field is assumed to be zero).
We will consider collective electrons in the metallic material. The model {\it jellium}\cite{brack,ekart,ksi}  is assumed in order to account for
screening background of  positive ions in the form of static uniformly distributed over the sphere positive charge:
\begin{equation}
n_e({\bm r})=n_e \Theta(a-r),
\end{equation}
 where  $n_e=N_e/V$  and $n_e|e|$ is the averaged positive charge density, $N_e$ the number of collective electrons in the sphere,
  $V=\frac{4 \pi a^3}{3}$ the sphere volume, and
$\Theta$ is the Heaviside step-function.  Neglecting the ion dynamics within  {\it jellium} model, which  is  adopted in particular for description
of simple metals, e.g. noble, transition and alkali metals, we deal with the Hamiltonian for collective electrons,
\begin{equation}
\label{h1}
 \hat{H}_{e} = \sum\limits_{j =1}^{N_e} \left[ -\frac{\hbar^2 \nabla_{j}^2}{2m} -e^2 \int\frac{n_e ({\bm r_0})d^3 {\bm r_0}}{|{\bm r}_{j}
   -  {\bm r_0}|} +e\varphi({\bm R}+{\bm r}_j, t)\right]
 +\frac{1}{2}\sum\limits_{j\neq j'}\frac{e^2}{|{\bm r}_{j}-{\bm r}_{j'}|} +\Delta E,
\end{equation}
where    ${\bm r}_{j}$ and $m$ are the position (with respect to the dot center) and the mass of the $j^{\text{th}}$ electron,
 ${\bm R}$ is the position of the metallic sphere center, $\Delta E$ represents  electrostatic
energy contribution from the ion 'jellium',
$ \varphi({\bm r}, t)$ is the scalar potential of the external electric field.
The corresponding electric field ${\bm E}({\bm R}+{\bm r}_j, t)=-grad_j\varphi({\bm R}+{\bm r}_j, t)$. Assuming that space-dependent variation
of ${\bm E}$ is weak on the scale of the sphere radius $a$ then
${\bm E}({\bm R}+{\bm r}_j, t) \simeq {\bm E}({\bm R},t)$, i.e the electric field is homogeneous over the sphere (it holds for $|{\bm R}|\gg a$).
Then $\varphi({\bm R}+{\bm r}_j, t)\simeq
-{\bm E}( {\bm R},t)\cdot  {\bm R} +  \varphi_1({\bm R}+{\bm r}_j, t)$, where
$\varphi_1({\bm R}+{\bm r}_j, t)=-{\bm r}_j\cdot\bm E( {\bm R},t)$.
Hence, one can rewrite the Hamiltonian (\ref{h1}) in the form:
\begin{equation}
\hat{H}_e=\hat{H}'_e - eN {\bm E}( {\bm R},t)\cdot  {\bm R},
 \end{equation}
 where
\begin{equation}
\label{h'}
 \hat{H}'_{e} = \sum\limits_{j =1}^{N_e} \left[ -\frac{\hbar^2 \nabla_{j}^2}{2m} -e^2 \int\frac{n_e ({\bm r})d^3 {\bm r_0}}{|{\bm r}_{j}
   -  {\bm r_0}|} +e\varphi({\bm r}_j, t)\right]
 +\frac{1}{2}\sum\limits_{j\neq j'}\frac{e^2}{|{\bm r}_{j}-{\bm r}_{j'}|} +\Delta E,
\end{equation}
the corresponding wave function can be represented as,
\begin{equation}
\Psi({\bm r_e},t)= \Psi'({\bm r_e},t)e^{i\frac{e N}{\hbar}\int  {\bm E}( {\bm R},t) \cdot {\bm R} dt }
\end{equation}
with $i\hbar \frac {\partial \Psi'}{\partial t}=\hat{H}'_{e}  \Psi'$, ${\bm r_e}=({\bm r_1},{\bm r_2},...,{\bm r_N})$.

A local electron density can be written as follows\cite{rpa}:
\begin{equation}
\rho({\bm r}, t)=<\Psi({\bm r_e},t)|\sum\limits_j \delta({\bm r}-{\bm r}_j) |\Psi({\bm r_e},t)>
= <\Psi'({\bm r_e},t)|\sum\limits_j \delta({\bm r}-{\bm r}_j) |\Psi'({\bm r_e},t)>,
\end{equation}
with the Fourier picture:
\begin{equation}
\tilde{\rho}({\bm k}, t)=\int \rho({\bm r},t) e^{-i{\bm k}\cdot {\bm r}} d^3 r = <\Psi'({\bm r_e},t)|\hat{\rho}({\bm k})|\Psi'({\bm r_e},t)>,
\end{equation}
where the 'operator' $ \hat{\rho} ({\bm k})=\sum\limits_{j}  e^{-i{\bm k}\cdot {\bm r_j}} $.

Using the above notation one can rewrite $\hat{H}'_{e}$ in the following form, in analogy to the bulk case\cite{pines}:
\begin{equation}
\begin{array}{l}
\hat{H}'_{e} = \sum\limits_{j =1}^{N_e} \left[ -\frac{\hbar^2 \nabla_{j}^2}{2m}\right]
-\frac{e^2}{4 \pi^2} \int d^3 k \tilde{n}_e({\bm k}) \frac{1}{k^2} \left(\hat{\rho^{+}}({\bm k}) +  \hat{\rho}({\bm k})\right) \\
 +\frac{e^2}{16 \pi^3} \int d^3 k \tilde{\varphi}_1({\bm k},t) \left(\hat{\rho^{+}}({\bm k})
  +  \hat{\rho}({\bm k})\right) +\frac{e^2}{4 \pi^2}\int d^3 k
  \frac{1}{k^2}\left[ \hat{\rho^{+}}({\bm k})  \hat{\rho}({\bm k}) -N_e \right]
  +\Delta E,\\
  \end{array}
\end{equation}
where:
$ \tilde{n}_e({\bm k})=\int d^3 r n_e ({\bm r}) e^{-i{\bm k}\cdot {\bm r}}$,
$  \frac{4 \pi}{k^2}= \int d^3 r \frac{1}{r} e^{-i{\bm k}\cdot {\bm r}}$,
    $ \tilde{\varphi}_1({\bm k})=\int d^3 r \varphi_1({\bm r},t) e^{-i{\bm k}\cdot {\bm r},t}$.

    Utilizing this form of the electron Hamiltonian one can write the secod time-derivative of  $ \hat{\rho} ({\bm k})$:
\begin{equation}
\frac{d^2 \hat{\rho} ({\bm k},t) }{dt^2}=\frac{1}{(i\hbar)^2} \left[\left[ \hat{\rho} ({\bm k}),\hat{H}'_{e} \right],\hat{H}'_{e} \right],
\end{equation}
which resolves itself into the equation:
\begin{equation}
\begin{array}{l}
\frac{d^2 \delta \hat{\rho} ({\bm k},t) }{dt^2}=-\sum\limits_{j}e^{-i{\bm k}\cdot {\bm r}_j}\left\{ -\frac{\hbar^2}{m^2}\left(
{\bm k}\cdot \nabla_j \right)^2
+ \frac{\hbar^2 k^2}{m^2}i  {\bm k}\cdot \nabla_j +\frac{\hbar^2 k^4}{4 m^2}\right\}\\
-\frac{ e^2}{m 2\pi^2}\int d^3q  \tilde{n}_e ({\bm k}-{\bm q})\frac{{\bm k}\cdot {\bm q}}{q^2} \delta \hat{\rho}( {\bm q})
-\frac{ e}{m 8\pi^3}\int d^3q  \tilde{n}_e ({\bm k}-{\bm q})({\bm k}\cdot {\bm q})\tilde{\varphi}_1({\bm q},t)\\
-\frac{ e}{m 8\pi^3}\int d^3q  \delta\hat{\rho} ({\bm k}-{\bm q})({\bm k}\cdot {\bm q})\tilde{\varphi}_1({\bm q},t)
-\frac{e^2}{m 2 \pi^2}\int d^3q \delta \hat{\rho}({\bm k}- {\bm q})\frac{{\bm k}\cdot {\bm q}}{q^2}  \delta \hat{\rho}( {\bm q}),\\
\end{array}
\end{equation}
where    $\delta \hat{\rho}({\bm k}) =  \hat{\rho}({\bm k)} -  \tilde{n}_e ({\bm k})$ is the 'operator' of
 local electron density fluctuations beyond the uniform distribution.
Taking into account that:
$\delta \tilde{\rho}({\bm k},t)=   <\Psi'(t)|\delta\hat{\rho}({\bm k})|\Psi'(t)>=  \tilde{\rho}({\bm k},t) -  \tilde{n}_e ({\bm k})$
we find:
\begin{equation}
\label{e10}
\begin{array}{l}
\frac{\partial^2 \delta \tilde{\rho} ({\bm k},t) }{\partial t^2}=<\Psi'|-\sum\limits_{j}e^{-i{\bm k}\cdot {\bm r}_j}\left\{ -\frac{\hbar^2}{m^2}\left(
{\bm k}\cdot \nabla_j \right)^2
+ \frac{\hbar^2 k^2}{m^2}i  {\bm k}\cdot \nabla_j +\frac{\hbar^2 k^4}{4 m^2}\right\}|\Psi'>\\
-\frac{ e^2}{m 2\pi^2}\int d^3q  \tilde{n}_e ({\bm k}-{\bm q})\frac{{\bm k}\cdot {\bm q}}{q^2} \delta \tilde{\rho}( {\bm q},t)
-\frac{ e}{m 8\pi^3}\int d^3q  \tilde{n}_e ({\bm k}-{\bm q})({\bm k}\cdot {\bm q})\tilde{\varphi}_1({\bm q},t)\\
-\frac{ e}{m 8\pi^3}\int d^3q  \delta\tilde{\rho} ({\bm k}-{\bm q},t)({\bm k}\cdot {\bm q})\tilde{\varphi}_1({\bm q},t)
-\frac{e^2}{m 2 \pi^2}\int d^3q\frac{{\bm k}\cdot {\bm q}}{q^2} <\Psi'| \delta \hat{\rho}({\bm k}- {\bm q}) \delta \hat{\rho}( {\bm q})|\Psi'>,\\
\end{array}
\end{equation}
One can simplify the above equation upon the assumption that $   \delta\rho({\bm r,t})=\frac{1}{8\pi^3}\int e^{i{\bm k}\cdot{\bm r}}
\delta \tilde{\rho}({\bm k},t)d^3 k$ only weakly varies on the interatomic scale, and hence
three components of the first term in right-hand-side of Eq. (\ref{e10}) can be estimated as:
 $k^2 v_F^2  \delta \tilde{\rho}({\bm k})$, $k^3 v_F/k_T\delta \tilde{\rho}({\bm k})$  and
 $ k^4 v_F^2/k_T^2\delta \tilde{\rho}({\bm k})$, respectively, with $1/k_T$ the Thomas-Fermi radius\cite{rpa},
 $k_T=\sqrt{\frac{6\pi n_e e^2}{\epsilon_F}}$,
 $\epsilon_F$ the Fermi energy, and $v_F$ the Fermi velocity.
 Thus the contribution of the second and the third components of the first term can be neglected in comparison to the first component.
 Small and thus negligible  is also
the  last term  in right-hand-side of Eq. (\ref{e10}), as it involves a product of two $ \delta \tilde{\rho}$ (which we
 assumed small\ $ \delta \tilde{\rho}/n_e << 1$).
This approach corresponds to random-phase-approximation (RPA) attitude formulated for bulk metal\cite{rpa,pines}
(note that $\delta \hat{\rho}(0)=0$ and the coherent RPA contribution
of interaction is comprised by the second   term in Eq. (\ref{e10})). The last but one term  in Eq. (\ref{e10})
can also be omitted if one confines it to linear terms with respect to $\delta \tilde{\rho}$ and $\tilde{\varphi_1}$.
Next, due to spherical symmetry,  $ <\Psi'|\sum\limits_{j}e^{-i{\bm k}\cdot {\bm r}_j}\frac{\hbar^2}{m^2}\left( {\bm k}\cdot \nabla_j \right)^2|\Psi'>
\simeq \frac{2 k^2}{3m}<\Psi'|\sum\limits_{j}e^{-i{\bm k}\cdot {\bm r}_j}\frac{\hbar^2\nabla_j^2}{2m}|\Psi'> $.
Performing the inverse Fourier transform,   Eq. (\ref{e10})  attains finally the form:
\begin{equation}
\label{e12}
\begin{array}{l}
\frac{\partial^2 \delta \rho ({\bm r},t) }{\partial ,t^2}=-\frac{2 }{3m} \nabla^2
<\Psi'|\sum\limits_{j}\delta({\bm r}-{\bm r}_j)\frac{\hbar^2\nabla_j^2}{2m}|\Psi'>\\
+\frac{\omega_p^2}{4\pi} \nabla \left\{ \Theta(a-r) \nabla \int d^3r_1 \frac{1}{|{\bm r}-{\bm r}_1|} \delta \rho( {\bm r}_1,t)\right\}
+\frac{en_e}{m} \nabla \left\{ \Theta(a-r) \nabla \varphi_1( {\bm r}_1,t)\right\}.\\
\end{array}
\end{equation}

According to Thomas-Fermi approximation\cite{rpa} the RPA  averaged kinetic energy can be represented as follows:
\begin{equation}
<\Psi'|-\sum\limits_{j}\delta({\bm r}-{\bm r}_j)\frac{\hbar^2\nabla_j^2}{2m}|\Psi'>\simeq
\frac{3}{5} (3\pi^2)^{2/3} \frac{\hbar^2}{2m} \rho^{5/3}({\bm r},t)=
\frac{3}{5} (3\pi^2)^{2/3} \frac{\hbar^2}{2m}n_e^{5/3} \Theta(a-r)\left[1+\frac{5}{3}\frac{\delta \rho({\bm r},t)}{n_e}+...\right].
\end{equation}
Taking then into account the above approximation and   that $\nabla \Theta(a-r)=-\frac{\bm r}{r}\delta(a-r)= -\frac{\bm r}{r}\lim_{\epsilon\rightarrow 0}\delta(a+\epsilon-r)$
as well as  that $\varphi_1({\bm R}, {\bm r}, t)=-{\bm r}\cdot {\bm E}({\bm R},t)$,
one can rewrite Eq. (\ref{e12}) in the following manner:
\begin{equation}
\label{e15}
\begin{array}{l}
\frac{\partial^2 \delta \rho ({\bm r}) }{\partial t^2}=\left[ \frac{2}{3} \frac{\epsilon_F}{m}\nabla^2 \delta \rho( {\bm r},t)-
\omega_p^2 \delta \rho( {\bm r},t)\right]\Theta(a-r)\\
- \frac{2}{3m} \nabla\left\{\left[\frac{3}{5}\epsilon_F n_e+\epsilon_F \delta \rho( {\bm r},t)\right]\frac{\bm r}{r}\delta(a+\epsilon-r)
\right\}-\left[\frac{2}{3} \frac{\epsilon_F}{m}\frac{\bm r}{r}\nabla \delta \rho( {\bm r},t)
+      \frac{\omega_p^2}{4\pi}          \frac{\bm r}{r}\nabla \int d^3r_1 \frac{1}{|{\bm r}-{\bm r}_1|} \delta \rho( {\bm r}_1,t)
+\frac{en_e}{m} \frac{\bm r}{r}\cdot{\bm E}({\bm R},t)\right]
\delta(a+\epsilon-r).\\
\end{array}
\end{equation}
In the above formula  $\omega_p$ is the bulk plasmon frequency,  $\omega_p^2=\frac{4\pi n_e e^2}{m}$, and
       $\delta(a+\epsilon-r)   =  \lim_{\epsilon\rightarrow 0}\delta(a+\epsilon-r)$.
The solution of Eq. (\ref{e15}) can be decomposed into two parts related  to  the domain:
\begin{equation}
 \delta \rho( {\bm r,t})=\left\{
           \begin{array}{l}
             \delta \rho_1( {\bm r,t}), \;for\; r<a,\\
              \delta \rho_2( {\bm r,t}), \;for\; r\geq a,\; ( r\rightarrow a+),\\
          \end{array}
       \right.
       \end{equation}
corresponding to the volume and surface  excitations, respectively. These two parts of local electron density fluctuations
satisfy the equations:
\begin{equation}
\label{e20}
\frac{\partial^2 \delta \rho_1 ({\bm r},t) }{\partial t^2}=\frac{2}{3} \frac{\epsilon_F}{m}\nabla^2 \delta \rho_1( {\bm r},t)-
\omega_p^2 \delta \rho_1( {\bm r},t),
\end{equation}
and
\begin{equation}
\label{e21}
\begin{array}{l}
\frac{\partial^2 \delta \rho_2 ({\bm r},t) }{\partial t^2} =-
\frac{2}{3m} \nabla\left\{\left[\frac{3}{5}\epsilon_F n_e+\epsilon_F \delta \rho_2( {\bm r},t)\right]\frac{\bm r}{r}\delta(a+\epsilon-r)\right\}\\
 -  \left[\frac{2}{3} \frac{\epsilon_F}{m}\frac{\bm r}{r}\nabla \delta \rho_2( {\bm r},t)
+      \frac{\omega_p^2}{4\pi}          \frac{\bm r}{r}\nabla \int d^3r_1 \frac{1}{|{\bm r}-{\bm r}_1|} \left(\delta \rho_1( {\bm r}_1,t)
\Theta(a-r_1)+\delta \rho_2( {\bm r}_1,t)\Theta(r_1-a)\right)+\frac{en_e}{m} \frac{\bm r}{r}\cdot{\bm E}({\bm R},t)\right]\delta(a+\epsilon-r).\\
\end{array}
\end{equation}
It is clear from Eq. (\ref{e20}) that the volume plasmons are independent of surface plasmons.
 However, surface plasmons can be excited by volume plasmons due to the last term in Eq. (\ref{e21}), which expresses a
  coupling between surface and volume plasmons
in the metallic nanosphere within RPA semiclassical picture. It is in fact a surface tail of volume compressional-type excitations, while surface traslational-type exctations
have no a volume tail.

In a dielectric medium in which the metallic sphere can be embedded, the electrons on the surface interact with
forces $\varepsilon$ (dielectric susceptibility constant) times  weaker in comparison to electrons inside the sphere. To account for it,
one substitutes Eqs (\ref{e20}) and (\ref{e21}) with the following ones:
\begin{equation}
\label{e200}
\frac{\partial^2 \delta \rho_1 ({\bm r},t) }{\partial t^2}=\frac{2}{3} \frac{\epsilon_F}{m}\nabla^2 \delta \rho_1( {\bm r},t)-
\omega_p^2 \delta \rho_1( {\bm r},t),
\end{equation}
and
\begin{equation}
\label{e210}
\begin{array}{l}
\frac{\partial^2 \delta \rho_2 ({\bm r},t) }{\partial t^2} =-
\frac{2}{3m} \nabla\left\{\left[\frac{3}{5}\epsilon_F n_e+\epsilon_F \delta \rho_2( {\bm r},t)\right]\frac{\bm r}{r}\delta(a+\epsilon-r)\right\}\\
 -  \left[\frac{2}{3} \frac{\epsilon_F}{m}\frac{\bm r}{r}\nabla \delta \rho_2( {\bm r},t)
+      \frac{\omega_p^2}{4\pi}          \frac{\bm r}{r}\nabla \int d^3r_1 \frac{1}{|{\bm r}-{\bm r}_1|} \left(\delta \rho_1( {\bm r}_1,t)
\Theta(a-r_1)+\frac{1}{\varepsilon}\delta \rho_2( {\bm r}_1,t)\Theta(r_1-a)\right)+\frac{en_e}{m} \frac{\bm r}{r}\cdot{\bm E}({\bm R},t)\right]
\delta(a+\epsilon-r).\\
\end{array}
\end{equation}

Let us also assume that both volume and surface plasmon oscillations are damped with the time ratio
$\tau_0$ which can be phenomenologically accounted for via the additional term,
$-\frac{2}{\tau_0}\frac{\partial \delta\rho({\bm r},t)}{\partial t}$, to the right-hand-side of above equations.
  They attain the form:
\begin{equation}
\label{e2000}
\frac{\partial^2 \delta \rho_1 ({\bm r},t) }{\partial t^2}+\frac{2}{\tau_0}\frac{\partial \delta\rho({\bm r},t)}{\partial t}=
\frac{2}{3} \frac{\epsilon_F}{m}\nabla^2 \delta \tilde{\rho}_1( {\bm r},t)-
\omega_p^2 \delta \rho_1( {\bm r},t),
\end{equation}
and
\begin{equation}
\label{e2100}
\begin{array}{l}
\frac{\partial^2 \delta \rho_2 ({\bm r},t) }{\partial t^2} +\frac{2}{\tau_0}\frac{\partial \delta\rho({\bm r},t)}{\partial t}=-
\frac{2}{3m} \nabla\left\{\left[\frac{3}{5}\epsilon_F n_e+\epsilon_F \delta \rho_2( {\bm r},t)\right]\frac{\bm r}{r}\delta(a+\epsilon-r)\right\}\\
 -  \left[\frac{2}{3} \frac{\epsilon_F}{m}\frac{\bm r}{r}\nabla \delta \tilde{\rho}_2( {\bm r},t)
+      \frac{\omega_p^2}{4\pi}          \frac{\bm r}{r}\nabla \int d^3r_1 \frac{1}{|{\bm r}-{\bm r}_1|} \left(\delta \rho_1( {\bm r}_1,t)
\Theta(a-r_1)+\frac{1}{\varepsilon}\delta \rho_2( {\bm r}_1,t)\Theta(r_1-a)\right)+\frac{en_e}{m}
 \frac{\bm r}{r}\cdot{\bm E}({\bm R},t)\right]\delta(a+\epsilon-r).\\
\end{array}
\end{equation}

From Eqs (\ref{e2000}) and (\ref{e2100}) it is noticeable that the homogeneous electric
field does not excite the volume-type plasmon oscillations but only contributes
to surface plasmons.

\subsection{Solution of RPA equations: volume and surface plasmons frequencies}

 Eqs (\ref{e2000}) and (\ref{e2100}) can be solved upon imposing the boundary and symmetry conditions---cf. Appendix \ref{app1}.
Let us write the both parts of the electron fluctuation in the following manner:
\begin{equation}
           \begin{array}{l}
             \delta \rho_1( {\bm r,t})=n_e\left[f_1(r)+F({\bm r}, t)\right], \;for\; r<a,\\
              \delta \rho_2( {\bm r,t})=n_e f_2(r)+\sigma(\Omega,t)\delta(r+\epsilon -a), \;for\; r\geq a,\; ( r\rightarrow a+),\\
          \end{array}
       \end{equation}
and let us choose the convenient initial conditions $  F({\bm r}, t)|_{t=0}=0, \; \sigma(\Omega,t)|_{t=0}=0$, ($\Omega=(\theta, \psi)$---the spherical angles), moreover
$(1+f_1(r))|_{r=a}=f_2(r)|_{r=a}$ (continuity condition), $ F({\bm r}, t)|_{r=a}=0$, $\int\rho({\bm r},t)d^3r=N_e$ (neutrality condition).

We thus arrive at the explicit form of the solutions of   Eqs (\ref{e2000}) and (\ref{e2100}) (as it is described in the Appendix \ref{app1}):
\begin{equation}
\label{fff}
\begin{array}{l}
f_1(r)=-\frac{k_T a +1}{2} e^{-k_T (a-r)} \frac{1-e^{-2k_Tr}}{k_Tr}, \; for \;\;r<a,\\
f_2(r)=\left[k_Ta - \frac{k_Ta+1}{2}\left(1-e^{-2k_Ta}\right)\right]\frac{e^{-k_T(r-a)}}{k_Tr}, \; for\;\; r\geq a,\\
\end{array}
\end{equation}
where $k_T=\sqrt{\frac{6\pi n_e e^2}{\epsilon_F}}=\sqrt{\frac{3\omega_p^2}{v_F^2}}$.
For time-dependent parts of electron fluctuations we find:
\begin{equation}
\label{e2001}
F({\bm r}, t) =\sum\limits_{l=1}^{\infty}\sum\limits_{m=-l}^{l}\sum\limits_{n=1}^{\infty}A_{lmn}j_{l}(k_{nl}r)Y_{lm}(\Omega)sin(\omega'_{nl}t)
e^{-t/\tau_0},
\end{equation}
and
\begin{equation}
\label{e25}
\begin{array}{l}
\sigma(\Omega,t)   = \sum\limits_{l=1}^{\infty}\sum\limits_{m=-l}^{l}Y_{lm}(\Omega)\left[ \frac{B_{lm}}{a^2}sin(\omega'_{0l}t)e^{-t/\tau_0}(1
-\delta_{1l})+ Q_{1m}(t) \delta_{1l}\right]\\
+ \sum\limits_{l=1}^{\infty}\sum\limits_{m=-l}^{l}\sum\limits_{n=1}^{\infty}
A_{lmn}\frac{(l+1)\omega_p^2}{l\omega_p^2-(2l+1)\omega_{nl}^2}Y_{lm}(\Omega)n_e\int\limits_0^a dr_1 \frac{r_1^{l+2}}{a^{l+2}}j_{l}(k_{nl}r_1)
sin(\omega'_{nl}t)e^{-t/\tau_0},\\
\end{array}
\end{equation}
where $j_l(\xi)=\sqrt{\frac{\pi}{2\xi}}I_{l+1/2}(\xi)$ is the spherical Bessel function, $Y_{lm}(\Omega)$ is the spherical function, $\omega_{nl}=
\omega_p\sqrt{1+\frac{x_{nl}^2}{k_T^2a^2}}$ are the frequencies of electron volume free self-oscillations (volume plasmon frequencies),
$x_{nl}$ are nodes of the Bessel function $j_l(\xi)$,
 $\omega_{0l}=\omega_p\sqrt{\frac{l}{\varepsilon (2l+1)}}$ are the frequencies of electron surface free self-oscillations (surface plasmon frequencies),
  and $k_{nl}=x_{nl}/a$; $\omega'=\sqrt{\omega^2-\frac{1}{\tau_0^2}}$ are the shifted frequencies for all   modes due to damping.
  The coefficients $ B_{lm}$ and  $A_{lmn}$ are determined by the initial conditions. As we have assumed that $\delta\rho({\bm r}, t=0)=0$,
  we get $ B_{lm}=0$ and  $A_{lmn}=0$, except for  $l=1$ in the former case (of $B_{lm}$) which corresponds to homogeneous electric
  field excitation. This  is described by the function $Q_{1m}(t)$ in the general solution (\ref{e25}).
 The  function $Q_{1m}(t)$ satisfies the equation:
  \begin{equation}
  \label{qqq}
  \begin{array}{l}
  \frac{\partial^2Q_{1m}(t)}{\partial t^2}+\frac{2}{\tau_0}\frac{\partial Q_{1m}(t)}{\partial t}+\omega_1^2 Q_{1m}(t)\\
   =\sqrt{\frac{4\pi}{3}}\frac{en_e}{m}\left[E_z({\bm R},t)\delta_{m0}+\sqrt{2}\left(E_x({\bm R},t)\delta_{m1}
   + E_y({\bm R},t)\delta_{m-1}\right)\right],\\
   \end{array}
   \end{equation}
   where  $\omega_1=\omega_{01}=\frac{\omega_p}{\sqrt{3\varepsilon}}$ (it is a dipole-type surface plasmon Mie frequency\cite{mie}).
   Only this function contributes the dynamical response to the homogeneous electric field (for the assumed  initial conditions).
From the above  it follows thus that  local electron density (within RPA attitude) has the form:
\begin{equation}
\label{e50}
\rho({\bm r},t)=\rho_0(r)+\rho_{1}({\bm r},t),
\end{equation}
where the  RPA  equilibrium electron distribution (correcting the uniform distribution $n_e$):
\begin{equation}
  \rho_0(r)=\left\{
  \begin{array}{l}
  n_e\left[1+f_1(r)\right],\; for\;\; r<a,\\
  n_ef_2(r),\;for\;\; r\geq a, \; r\rightarrow a+\\
  \end{array} \right.
  \end{equation}
and the nonequilibrium, of surface  plasmon oscillation type for the homogeneous forcing field:
\begin{equation}
\label{oscyl}
  \rho_{1}({\bm r},t)=\left\{
  \begin{array}{l}
  0,\; for\;\; r<a,\\
\sum\limits_{m=-1}^{1}Q_{1m}(t)Y_{1m}(\Omega)\;for\;\; r\geq a,\; r\rightarrow a+.\\
  \end{array} \right.
  \end{equation}
In general, $ F({\bm r}, t)$ (volume plasmons) and $\sigma(\Omega,t)$ (surface plasmons) contribute to plasmon oscillations. However,
in the case homogeneous perturbation, only the surface $l=1$ mode is excited.

For plasmon oscillations given by Eq. (\ref{oscyl}) one can calculate the corresponding dipole,
\begin{equation}
\label{dipol}
{\bm D}({\bm R},t)= e\int d^3r {\bm r}\rho({\bm r},t)=  \frac{4\pi}{3}e{\bm q}({\bm R},t)a^3,
\end{equation}
where
$Q_{11}({\bm R}, t)=\sqrt{\frac{8\pi}{3}}q_x({\bm R}, t)$,  $Q_{1-1}({\bm R}, t)=\sqrt{\frac{8\pi}{3}}q_y({\bm R}, t)$,
   $Q_{10}({\bm R}, t)=\sqrt{\frac{4\pi}{3}}q_x({\bm R}, t)$
   and ${\bm q}({\bm R},t)$ satisfies the equation (cf. Eq. (\ref{qqq})),
   \begin{equation}
   \label{dipoleq}
   \left[\frac{\partial^2}{\partial t^2}+  \frac{2}{\tau_0}  \frac{\partial}{\partial t} +\omega_1^2\right] {\bm q}({\bm R},t)=\frac{en_e}{m}
   {\bm E}({\bm R},t).
   \end{equation}

\section{Lorentz friction for nanosphere plasmons}

Considering the nanosphere plasmons induced by the homogeneous electric field, as described
in the above paragraph, one can note that these plasmons are themselves a source of the e-m radiation. This
radiation takes away the energy of plasmons resulting in their damping, which can be described as the Lorentz friction\cite{lan}.
This e-m wave emission causes electron  friction which  can be described as the additional electric field\cite{lan},
\begin {equation}
{\bm E}_L= \frac{2}{3\varepsilon v^2}\frac{\partial^3{\bm D}(t)}{\partial t^3},
\end{equation}
where $v=\frac{c}{\sqrt{\varepsilon}}$ is the light velocity in the dielectric medium, and ${\bm D}(t)$ the dipole of the nanosphere.
According to Eq. (\ref{dipol}) we arrive at the following:
\begin{equation}
\label{lor}
{\bm E}_L= \frac{2e}{3\varepsilon v^2}\frac{4\pi}{3}a^3\frac{\partial^3{\bm q}(t)}{\partial t^3}.
\end{equation}
Substituting it into Eq. (\ref{dipoleq}) we get
\begin{equation}
\left[\frac{\partial^2}{\partial t^2}+  \frac{2}{\tau_0}  \frac{\partial}{\partial t} +\omega_1^2\right] {\bm q}({\bm R},t)=\frac{en_e}{m}
   {\bm E}({\bm R},t) +\frac{2}{3\omega_1}\left(\frac{\omega_1a}{v}\right)^3\frac{\partial^3{\bm q}(t)}{\partial t^3}.
   \end{equation}
   If rewrite the above equation (for ${\bm E}$=0) in the form
   \begin{equation}
   \label{appr1}
\left[\frac{\partial^2}{\partial t^2} +\omega_1^2\right] {\bm q}({\bm R},t)=
 \frac{\partial}{\partial t}\left[ -\frac{2}{\tau_0} +
\frac{2}{3\omega_1}\left(\frac{\omega_1a}{v}\right)^3\frac{\partial^2{\bm q}(t)}{\partial t^2}\right],
   \end{equation}
 thus the zeroth order approximation (neglecting attenuation) corresponds to the equation;
 \begin{equation}
 \label{appr}
\left[\frac{\partial^2}{\partial t^2} +\omega_1^2\right] {\bm q}({\bm R},t)= 0.
\end{equation}
In order to solve Eq. (\ref{appr1}) in the next step of perturbation, in the right-hand-side of this equation one can
subsitute   $\frac{\partial^2{\bm q}(t)}{\partial t^2}$ by $-\omega_1^2 {\bm q}(t) $ (acc. to Eq. (\ref{appr})).

Therefore, if one assumes  the above estimation, $ \frac{\partial^3{\bm q}(t)}{\partial t^3}\simeq -\omega_1^2    \frac{\partial{\bm q}(t)}{\partial t}$,
then one can include the Lorentz friction in a  renormalised damping term:
\begin{equation}
\label{ratio}
  \left[\frac{\partial^2}{\partial t^2}+  \frac{2}{\tau}  \frac{\partial}{\partial t} +\omega_1^2\right] {\bm q}({\bm R},t)=\frac{en_e}{m}
   {\bm E}({\bm R},t) ,
   \end{equation}
   where
   \begin{equation}
   \label{tau}
   \frac{1}{\tau}=\frac{1}{\tau_0}+\frac{\omega_1}{3}\left(\frac{\omega_1 a}{v}\right)^3\simeq \frac{v_F}{2\lambda_B}+\frac{Cv_F}{2a}
   +  \frac{\omega_1}{3}\left(\frac{\omega_1 a}{v}\right)^3,
   \end{equation}
 where we used for $\frac{1}{\tau_0}\simeq  \frac{v_F}{2\lambda_B}+\frac{Cv_F}{2a} $
  ($\lambda_B$ is the free path in bulk, $ v_F$ the Fermi velocity, and $C\simeq 1$ a constant)\cite{atwater,kr}
 which corresponds to inclusion of plasmon damping due to electron scattering on other electrons and on nanoparticle boundary.
 The renormalised  damping causes the change in the shift of self-frequencies of free surface plasmons,
 $\omega_1'=\sqrt{\omega_1^2-\frac{1}{\tau^2}}$.

 Using Eq. (\ref{tau}) one can determine the radius $a_0$ corresponding to minimal damping,
 \begin{equation}
 \label{a0}
 a_0=\frac{\sqrt{3}}{\omega_p}\left(v_Fc^3\sqrt{\varepsilon}/2\right)^{1/4}.
 \end{equation}
For nanoparticles of gold, silver and copper in air, in water and in a colloidal solution, one can find
 $a_0 \leq 10$nm (cf. Tab. 1),
which corresponds to the experimental data\cite{ccc,stietz,scharte}. For $a>a_0$ damping increases due to Lorentz friction (proportional to $a^3$)
but for $a<a_0$ damping due to electron scattering dominates and causes also damping enhancement (with lowering $a$, as $\sim \frac{1}{a}$, cf. Fig. 1),

\begin{figure}[tb]
\centering
\scalebox{0.4}{\includegraphics{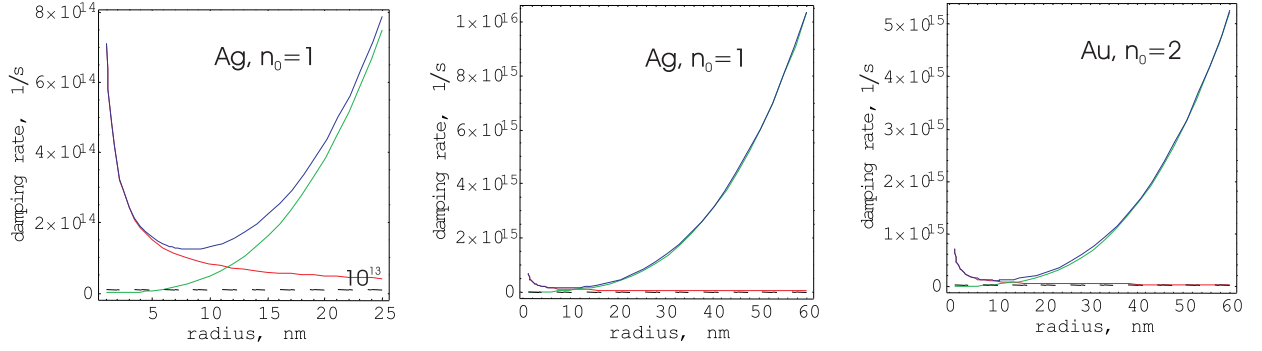}}
\caption{\label{fig1} Effective damping ratio for surface plasmon oscillations, Eqs (\ref{ratio}), (\ref{tau}), the upper (blue) curve is the sum
of both terms, $\sim \frac{1}{a}$ (red) and $\sim a^3$ (green); the minimum corresponds to minimal damping for radius $a_0$, Eq. (\ref{a0}), left---for Ag in the air,
 right---for Au in colloidal water solution}
\end{figure}
 which agrees with experimental observations\cite{atwater,ccc}.

\begin{tabular}{|p{8cm}|p{2.5cm}|p{2.5cm}|p{2.5cm}|}
\multicolumn{4}{c}{Tab. 1. $a_0$---nanosphere radius corresponding to minimal damping \label{tab1}}\\
\hline
refraction rate of the surrounding medium, $n_0$ & Au, $a_0$ [nm] & Ag, $a_0$ [nm]& Cu,  $a_0$ [nm] \\
\hline
(air) 1   & 8.8 &8.44&8.46\\
\hline
(water) 1.4   & 9.14&9.18&9.20\\
\hline
(colloidal solution) 2 &9.99& 10.04&10.04\\
\hline
\end{tabular}
 \vspace{5mm}

Surface plasmon oscillations cause attenuation of the incident e-m radiation where the maximum of attenuation is at the resonant frequency\cite{jac}
$\omega_1=\sqrt{\omega_1^2-\frac{1}{\tau^2}}$. This frequency diminishes with rise of $a$, for $a>a_0$ according to Eq. (\ref{tau}),
which agrees with experimental observations for Au and Ag presented in Fig. 2, and Tab. 2 (Au) and Tab 3 (Ag).

 \begin{figure}[tb]
\centering
\scalebox{0.8}{\includegraphics{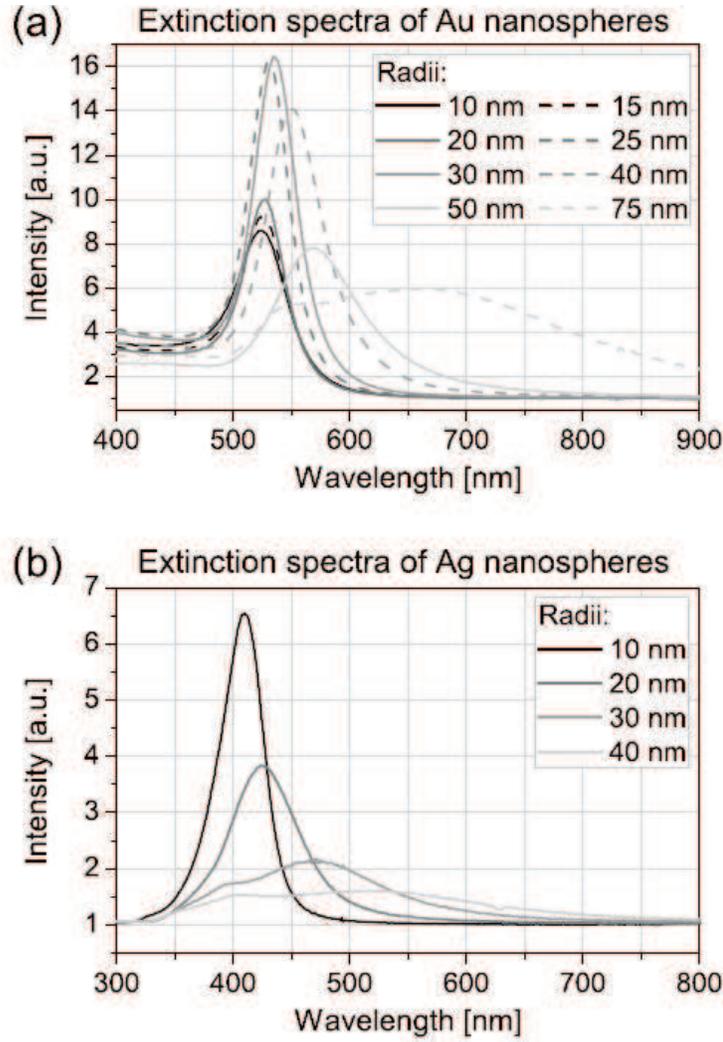}}
\caption{\label{fig2} Extinction spectra for nanospheres of Au (a) and Ag (b) in colloidal water solution for various sphere radii }
\end{figure}

\vspace{2mm}

\begin{tabular}{|p{4.5cm}|p{1.2cm}|p{1.2cm}|p{1.2cm}|p{1.2cm}|p{1.2cm}|p{1.2cm}|p{1.2cm}|}
\multicolumn{8}{c}{Tab. 2. Resonant frequency for e-m wave attenuation in Au nanospheres \label{tab2}}\\
\hline
radius of nanosheres [nm]                    & 10    & 15    & 20    & 25    & 30    & 40    & 50  \\
\hline
$\hbar\omega_1'$ (experiment) [eV]           & 2.371 & 2.362 & 2.357 & 2.340 & 2.316 & 2.248 & 2.172 \\
\hline
$\hbar\omega_1'$ (theory) [eV], $n_0= 1.4$   & 3.721 & 3.720  & 3.716 & 2.702& 3.666 & 3.415  & 2.374 \\
\hline
$\hbar\omega_1'$ (theory) [eV], $n_0= 2$     &  2.604 & 2.603 & 2.600 & 2.590 & 2.565 & 2.388 &1.656\\
\hline
\end{tabular}

\begin{tabular}{|p{4.5cm}|p{1.2cm}|p{1.2cm}|p{1.2cm}|p{1.2cm}|}
\multicolumn{5}{c}{Tab. 3. Resonant frequency for e-m wave attenuation in Ag nanospheres \label{tab3}}\\
\hline
radius of nanosheres [nm]                    & 10    & 20    & 30    & 40\\
\hline
$\hbar\omega_1'$ (experiment) [eV]           & 3.024 & 2.911 & 2.633 & 2.385\\
\hline
$\hbar\omega_1'$ (theory) [eV], $n_0= 1.4$   & 3.707 &  3.702&  3.654 & 3.410 \\
\hline
$\hbar\omega_1'$ (theory) [eV], $n_0= 2$     &  2.595 & 2.591 & 2.557 & 2.384\\
\hline
\end{tabular}

\section{Plasmon-mediated energy transfer through a chain of metallic nanospheres}

Let us consider a linear chain of metallic nanospheres with radii $a$ in a dielectric medium with dielectric constant $\varepsilon$.
We assume that spheres are located along $z$-axis direction equidistantly with separation of sphere centers $d>2a$\cite{atwater,atwater1}. At time $t=0$ we assume the
excitation of plasmon oscillation via a Dirac delta $\sim \delta(t)$ shape signal of electric field. Taking into account
the mutual interaction of induced surface plasmons on the spheres via the radiation of  dipole oscillations, we aim to determine
the stationary state of  the whole infinite chain. For separation $d$ much shorter than the wavelength $\lambda$ of the e-m wave corresponding to
surface plasmon self-frequency, the dipole type plasmon radiation
can be treated within near-field regime, at least for nearest neighbouring spheres. In the near-field region $a<R_0<\lambda$,
the radiation of the dipole ${\bm D}(t)$ is not a planar wave (as for far-field region, $R_0\gg \lambda$) but of
only electric field type in retarded form (without magnetic field)\cite{lan}:
\begin{equation}
{\bm E}({\bm R}, {\bm R}_0, t)=\frac{1}{\varepsilon R_0^3}\left[3{\bm n}\left({\bm n}\cdot {\bm D}\left({\bm R},
t-\frac{R_0}{v}\right)\right)- {\bm D}\left({\bm R}, t-\frac{R_0}{v}\right) \right],
\end{equation}
${\bm R}$---position of the sphere (center) irradiating e-m energy due to its dipole surface plasmon oscillations,  ${\bm R}_0$ position of
another sphere (center), with respect to the center of the former one, where the field ${\bm E}({\bm R}, {\bm R}_0, t)$ is given by the above formula, $R_0< \lambda$,
${\bm n}={\bm R}_0/R_0$, $v=c/\sqrt{\varepsilon}=c/n_0$.

When both vectors  ${\bm R}$ and ${\bm R}_0$ are along the $z$-axis (the linear chain) the above equation can be resolved as:
\begin{equation}
\label{daa}
 {\bm E}_{\alpha}({\bm R}, {\bm R}_0, t)=\frac{\sigma_\alpha}{\varepsilon R_0^3}D_{\alpha}\left({\bm R}, t-\frac{R_0}{v}\right),
\end{equation}
where $\alpha=(x,y,z)$, $\sigma_x=\sigma_y=-1$ and $\sigma_z=2$.
Assuming that the $z$-axis origin coincides with the center of one sphere in the chain, for the $l^{\text{th}}$ sphere located in the point
${\bm R}_l=(0,0,ld)$, an electric  field caused by neighbouring spheres,
${\bm E}({\bm R}_m, {\bm R}_{ml}, t)$, and the Lorentz friction force caused by self-radiation, ${\bm E}_L({\bm R}_l,t)$, has to be considered.
By virtue of  Eq. (\ref{dipoleq}) the equation for surface plasmon oscillation of the $l^{\text{th}}$ sphere is
\begin{equation}
\label{tra1}
 \left[\frac{\partial^2}{\partial t^2}+  \frac{2}{\tau_0}  \frac{\partial}{\partial t} +\omega_1^2\right] {\bm q}({\bm R}_l,t)=\frac{en_e}{m}
  \sum\limits_{m=-\infty, \;m\ne l, \; R_{ml}<\lambda}^{m=\infty} {\bm E}({\bm R}_m, {\bm R}_{ml},t)+\frac{en_e}{m}{\bm E}_L({\bm R}_l,t),
\end{equation}
provided that the dipole field of the $m^{\text{th}}$ sphere can be treated as homogeneous over the $l^{\text{th}}$ sphere and  the sum over $m$
is confined by the distance of $m^{\text{th}}$ sphere from $l^{\text{th}}$ sphere not exceeding the near-field range ($\sim \lambda$).
In the case of the equidistant chain, $R_l=ld$ and $R_{ml}=|l-m|d$, and using Eqs (\ref{daa}), (\ref{dipol}) and  {\ref{lor}),
one can rewrite Eq. (\ref{tra1}) in the form:
\begin{equation}
\label{qequat}
  \left[\frac{\partial^2}{\partial t^2}+  \frac{2}{\tau_0}  \frac{\partial}{\partial t}-\frac{2}{3\omega_1}\left(\frac{\omega_1a}{v}\right)^3
  \frac{\partial^3}{\partial t^3} +\omega_1^2\right]  q_{\alpha}(ld,t)=\sigma_{\alpha}\frac{a^3}{d^3}
  \sum\limits_{m=-\infty, \;m\ne l, \; |l-m|d<\lambda}^{m=\infty}\frac{q_{\alpha}\left(md, t-\frac{d}{v}|l-m|\right)}{|l-m|^3},
  \end{equation}
  here $\alpha=x,y$, which describe the transversal plasmon modes and $\alpha=z$, which describes the longitudinal one.
 The above equation coincides with the appropriate one from Refs [\onlinecite{atwater,atwater1}], if one assumes  that $\frac{4\pi}{3}a^3n_e=N=1$ and
 neglects the retardation of the field.

 Taking into account the periodicity of the infinite chain, one can consider the solution of the above equation in the form
 \begin{equation}
 q(ld,t)=\tilde{q}(k,t)e^{-ikld}.
 \end{equation}
The right-hand-side term in Eq. (\ref{qequat}) attains the form
 $$
 \begin{array}{l}
 \sum\limits_{m=-\infty, \;m\ne l}^{m=\infty}\frac{q_{\alpha}\left(md, t-\frac{d}{v}|l-m|\right)}{|l-m|^3}
 = \sum\limits_{m=-\infty}^{l-1}\frac{q_{\alpha}\left(md, t-\frac{d}{v}|l-m|\right)}{|l-m|^3}
+  \sum\limits_{m=l+1}^{m=\infty}\frac{q_{\alpha}\left(md, t-\frac{d}{v}|l-m|\right)}{|l-m|^3}\\
=
2e^{-ikld}\sum\limits_{m=1}^{\infty}\frac{cos(mkd)}{m^3}\tilde{q}(k,t-md/v).\\
\end{array}
$$
Thus the Eq. (\ref{qequat}) can be written as follows:
\begin{equation}
  \left[\frac{\partial^2}{\partial t^2}+  \frac{2}{\tau_0}  \frac{\partial}{\partial t}-\frac{2}{3\omega_1}\left(\frac{\omega_1a}{v}\right)^3
  \frac{\partial^3}{\partial t^3} +\omega_1^2\right] \tilde{q}_{\alpha}(k,t)=\sigma_{\alpha}\omega_1\frac{a^3}{d^3}
  \sum\limits_{m=1,\; md<\lambda}^{\infty}\frac{cos(mkd)}{m^3}\tilde{q}(k,t-md/v)
\end{equation}
 This equation is linear and therefore  we look for the solutions  of the shape:
  $\tilde{q}_{\alpha}(k,t)=\tilde{Q}_{\alpha}(k)e^{i\omega_{\alpha}t}$, and
  \begin{equation}
  \label{omeg}
  -\omega_{\alpha}^2+\frac{2i\omega_{\alpha}}{\tau_{\alpha}(\omega_{\alpha})}+\tilde{\omega}_{\alpha}^2(\omega_{\alpha})=0,
  \end{equation}
where
\begin{equation}
\label{omega}
 \tilde{\omega}_{\alpha}^2(\omega_{\alpha})=\omega_1\left[1-\frac{2\sigma_{\alpha}a^3}{d^3}
         \sum\limits_{m=1,\; md<\lambda}^{\infty}\frac{cos(mkd)}{m^3} cos\left(\frac{\omega_{\alpha}md}{v}\right)\right]
         \end{equation}
and
\begin{equation}
\label{taue}
      \frac{1}{\tau_{\alpha}(\omega_{\alpha})}=  \frac{1}{\tau_0}+\frac{\omega_1^2a}{3v}\left(\frac{\omega_{\alpha}a}{v} \right)^2
        +\sigma_{\alpha}\omega_1^2\frac{a^3}{d^3}   \sum\limits_{m=1,\; md<\lambda}^{\infty}\frac{cos(mkd)}{m^3}
                                        \frac{sin\left(\frac{\omega_{\alpha}md}{v}\right)}{\omega_{\alpha}}.
                                        \end{equation}

If we confine the sum in the Eq. (\ref{omega}) to $m=1$ (the nearest neighbour approximation) we get
\begin{equation}
    \tilde{\omega}_{\alpha}^2(\omega_{\alpha})\simeq\omega_1\left[1-\frac{2\sigma_{\alpha}a^3}{d^3}
        cos(kd) cos\left(\frac{\omega_{\alpha}d}{v}\right)\right]
         \end{equation}
and from Eq. (\ref{taue}),
\begin{equation}
   \frac{1}{\tau_{\alpha}(\omega_{\alpha})}=  \frac{1}{\tau_0}+\frac{\omega_1^2a^3}{4vd^2}\left[\left(\frac{\omega_{\alpha}d}{v}\right)^2
   -(kd-\pi)^2+\frac{\pi^2}{3}\right], \;\; for\;\alpha=x,y
   \end{equation}
   and
\begin{equation}
   \frac{1}{\tau_{z}(\omega_z)}=  \frac{1}{\tau_0}+\frac{\omega_1^2a^3}{2vd^2}\left[\left(\frac{\omega_zd}{v}\right)^2
   +(kd-\pi)^2-\frac{\pi^2}{3}\right], \;\; for\;\alpha=z .
   \end{equation}

                                                           In the derivation of two above formulae the following
                                                           summation was performed\cite{grad}:
                                                           $$
                                                           \begin{array}{l}
                                                          \frac{1}{\omega_{\alpha}}  \sum\limits_{m=1}^{\infty}\frac{cos(mkd)}{m^3}
                                       sin\left(\frac{\omega_{\alpha}md}{v}\right)=
     \frac{1}{2\omega_{\alpha} } \sum\limits_{m=1}^{\infty} \frac{1}{m^3}[sin(kmd+\omega_{\alpha}md/v)-sin(kmd-\omega_{\alpha}md/v)]\\
     =\frac{d}{v}\left[\frac{\pi^2}{6}-\frac{\pi}{2}kd+\frac{k^2d^2}{4}+\frac{\omega_{\alpha}^2d^2}{2v^2}\right],\\
     \end{array}
     $$
as the terms in the sum drop quickly to zero then the above formula well approximates the sum with limitation $md<\lambda$.

Assuming now $\omega_{\alpha}= \omega_{\alpha}^{'}  +i \omega_{\alpha}^{''}$ the Eq. (\ref{omeg})
gives the dependence of $\omega_{\alpha}^{'}$ and $\omega_{\alpha}^{''}$ on $k$.
The general solution of Eq. (\ref{qequat}) attains the form,
\begin{equation}
\label{dae}
q_{\alpha}(ld,t)=\sum\limits_{n=1}^{N_s}\tilde{Q}_{\alpha}(k_n)e^{i(\omega_{\alpha}^{'}(k_n)t-k_nld)-\omega_{\alpha}^{''}(k_n)t},
\end{equation}
where $k_n=\frac{2\pi n}{N_sd}$, $L=N_s d$ is assumed length of the chain with $N_s$ spheres, and periodic (of Born-Karman type)
boundary condition imposed.
The components of Eq. (\ref{dae}) describe monochromatic  waves with wavelength $\lambda_n=\frac{2\pi}{k_n}=\frac{L}{n}$, which are
 analogous to planar waves
in crystals, when damping is not big, i.e., when $\omega_{\alpha}^{''}\ll \omega_{\alpha}^{'}$. Provided this inequality one can approximate:

for transversal modes ($\alpha=x,y$)

\begin{equation}
\label{333}
(\omega_{\alpha}^{'})^2 =(\tilde{\omega}_{\alpha})^2=\omega_1^2\left[1+\frac{2a^3}{d^3}cos(kd)cos(\omega_{\alpha}^{'}d/v)\right],
\end{equation}
\begin{equation}
\label{111}
\omega_{\alpha}^{''}=\frac{1}{\tau_{\alpha}}=\frac{1}{\tau_0}+\frac{\omega_1^2a^3}{4vd^2}\left[\left(\frac{\omega_{\alpha}^{'}d}{v}\right)^2
-(kd-\pi)^2+\frac{\pi^2}{3}\right],
\end{equation}

and for longitudinal mode    ($\alpha=z$)

\begin{equation}
\label{666}
(\omega_z^{'})^2 =(\tilde{\omega}_z)^2=\omega_1^2\left[1-\frac{4a^3}{d^3}cos(kd)cos(\omega_z^{'}d/v)\right],
\end{equation}
\begin{equation}
\label{222}
\omega_z^{''}=\frac{1}{\tau_z}=\frac{1}{\tau_0}+\frac{\omega_1^2a^3}{2vd^2}\left[\left(\frac{\omega_{\alpha}^{'}d}{v}\right)^2
+(kd-\pi)^2-\frac{\pi^2}{3}\right].
\end{equation}

 From the Eqs (\ref{111}) and (\ref{222}) it follows that $\omega_{\alpha}^{''}$ can change its sign. In the case of $\omega_{\alpha}^{''} < 0$
 the oscillations are destabilized, which could be avoided by inclusions of some nonlinear terms
 neglected in the expression for the Lorentz friction, which in  more accurate form\cite{lan}
includes also a small nonlinear term with respect to $D$, aside from the term with $\frac{\partial^3 D}{\partial t^3}$. Including of it will result in damping of too highly rising oscillations leading to stable amplitude of oscillations.
Due to  this stabilisation caused by nonlinear effects, undamped wave modes
of dipole oscillations will propagate in the chain in the region of parameters  where $   \omega_{\alpha}^{''}\leq 0$ (and with fixed amplitude accommodated by nonlinear term).
 The condition $   \omega_{\alpha}^{''}=\frac{1}{\tau_{\alpha}}= 0$, for critical parameters,  resolves into:
 \begin{equation}
 \left(\frac{\omega_{\alpha}d}{v}\right)^2=(kd-\pi)^2-\frac{\pi^2}{3}-\frac{4vd^2}{\tau_0\omega_1^2a^3},
 \end{equation}
 for $\alpha=(x,y)$ and for $\alpha =z $,
  \begin{equation}
 \left(\frac{\omega_zd}{v}\right)^2=-(kd-\pi)^2+\frac{\pi^2}{3}-\frac{2vd^2}{\tau_0\omega_1^2a^3}.
 \end{equation}
  Obtained from the above equations  $ \left(\frac{\omega_{\alpha}d}{v}\right)$ leads to determination of the dependence of wave vector
  $k$ with respect to parameters $d$ and $a$, via Eqs (\ref{333})-(\ref{222}). Solution for this equations, found numerically  for the chain of Ag nanospheres,
   is depicted in the Fig. 3.

 \begin{figure}[tb]
\centering
\scalebox{0.4}{\includegraphics{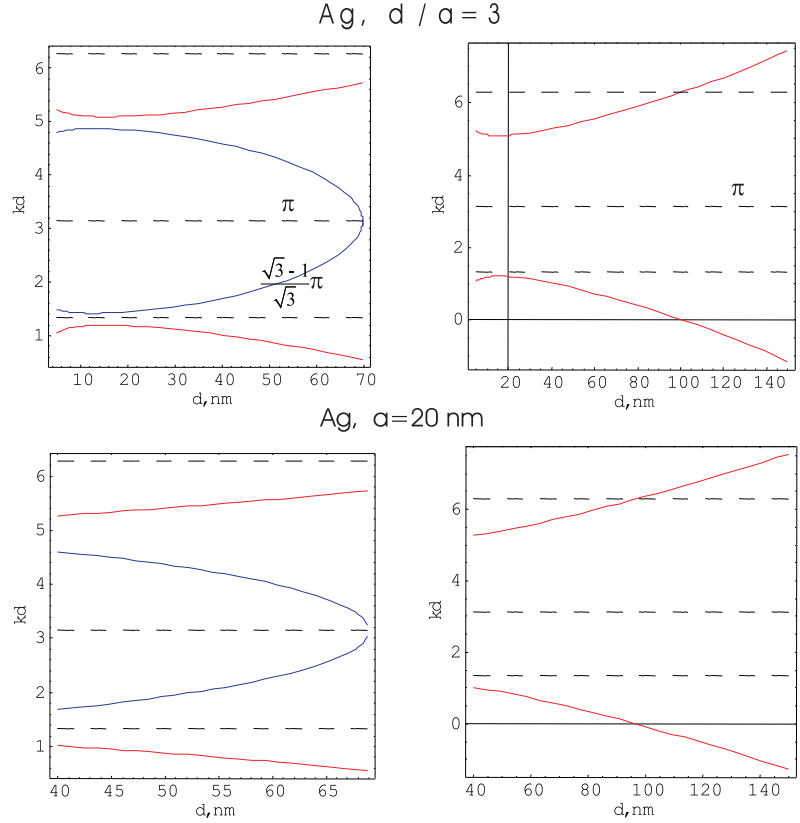}}
\caption{\label{fig3} Wave vector $kd$ versus sphere separation $d$, at constant $d/a=3$ ($a$---sphere radius) (upper),
for $a=20$ nm (lower) for Ag sphere chain, transversal modes---red, longitudinal mode---blue }
\end{figure}

   The undamped plasmon waves in the chain appear if $d<d_{max}$ and have $k=\pi/d$, $d_{max}=98.5(68.8)$ nm for transversal(longitudinal) modes.
   For example, for Ag spheres with radius $a=20$nm and separation $d=60$ nm, the undamped transversal modes appear for
     $0\leq kd\leq \pi/4$ or $3\pi/4\leq kd\leq 2\pi$ and longitudinal for $3\pi/4\leq kd \leq 5\pi/4$.

     Let us underline that the determined undamped plasmon oscillation wave modes explain the numerically observed
     similar behaviour\cite{ggg}.

\section{Conclusions}

In the present paper we analysed plasmons in large metallic nanospheres induced by homogeneous time-dependent electric field.
Within all-analytical RPA quasiclassical approach the volume and surface plasmons are described and a proof
 that only dipole-type of surface plasmons
can be induced by a homogeneous field (while none of volume modes) is given. An irradiation of energy by plasmon oscillations is
described within the Lorentz friction effect. Its scaling with the nanosphere dimension leads to sphere radius dependent shift
of resonant frequency, similarly as observed in experiments. The description of surface dipole-type plasmon
oscillations in single nanospheres is applied to analysis of collective oscillation in linear chain of metallic nanospheres.
The wave-type collective plasmon oscillations in the chain are also considered. The undamped region of wave energy transport through the chain is
found for a certain sphere separation in the chain with corresponding appropriate wavelength of plasmon waves. This phenomenon
confirms a  similar behaviour observed by numerical simulations\cite{ggg}.

 \begin{acknowledgments}
 Supported by the Polish KBN Project No:  N N202 260734 and the FNP Fellowship Start (W. J.), as well as DFG grant SCHA 1576/1-1 (D. S. and D. Z. Hu)

\end{acknowledgments}

\appendix

   \section{Analytical solution of plasmon equations for the nanosphere }

\label{app1}

Let us solve first the Eq. (\ref{e2000}), assuming a solution in the form:
\begin{equation}
 \delta\rho_1( {\bm r,t})=n_e\left[f_1(r)+F({\bm r}, t)\right], \;for\; r<a,
 \end{equation}
    Eq. (\ref{e2000}) resolves thus into:
\begin{equation}
\label{ea1}
\begin{array}{l}
\nabla^2 f_1(r)-k_T^2  f_1(r) =0,\\
\frac{\partial^2 F({\bm r},t)}{\partial t^2}+\frac{2}{\tau_0}\frac{\partial F({\bm r},t)}{\partial t}=\frac{v_F^2}{3}\nabla^2 F({\bm r},t) -\omega_p^2  F({\bm r},t),\\
\end{array}
\end{equation}
The solution for function $f_1(r)$ (nonsingular at $r=0$) has thus the form:
\begin{equation}
\label{alpha}
f_1(r)=\alpha \frac{e^{-k_Ta}}{k_Tr}\left(e^{-k_Tr}  -  e^{k_Tr}\right),
\end{equation}
where $\alpha$---const.,  $k_T=\sqrt{\frac{6\pi n_e e^2}{\epsilon_F}}=\sqrt{\frac{3\omega_p^2}{v_F^2}}\;$     ($k_T$---inverse  Thomas-Fermi radius),
$\omega_p=\sqrt{\frac{4\pi n_e e^2}{m}}\;$ (bulk plasmon frequency).

Since we assume
 $ F({\bm r},0)=0$, then for function     $ F({\bm r},t)$  the solution can be taken as,
 \begin{equation}
 F({\bm r}, t)= F_{\omega}({\bm r}) sin(\omega' t)e^{-\tau_0t}
 \end{equation}
 where
 $\omega'=\sqrt{\omega^2+ 1/\tau_0^2}$.
$F_{\omega}({\bm r})$ satisfies the equation (Helmholtz equation):
\begin{equation}
\nabla^2  F_{\omega}({\bm r})+k^2 F_{\omega}({\bm r})=0,
\end{equation}
with $k^2=\frac{\omega^2-\omega_p^2}{v_F^2/3}$.
A solution of the above equation, nonsingular at $r=0$, is as follows:
\begin{equation}
 F_{\omega}({\bm r})=Aj_l(kr)Y_{lm}(\Omega),
 \end{equation}
 where $A$---constant, $j_l(\xi)=\sqrt{\pi/(2\xi)}I_{l+1/2}(\xi)$---the spherical Bessel function
  [$I_{n}(\xi)$---the Bessel function of the first order],
 $Y_{lm}(\Omega)$---the spherical function ($ \Omega$---the spherical angle).
Owing to the boundary condition,   $ F({\bm r}, t)|_{r=a}=0$, one has to demand $j_l(ka)=0$, which leads to the discrete values of $k=k_{nl}=x_{nl}/a$,
(where $x_{nl},\;\;n=1,2,3...$, are nodes of $j_{l}$), and next to the discretisation of self-frequencies $\omega$:
\begin{equation}
\omega_{nl}^2=\omega_{p}^2\left(1+\frac{x_{nl}^2}{k_T^2a^2}\right).
\end{equation}

 The general solution for $F({\bm r},t)$ has thus the form
\begin{equation}
\label{ea10}
F({\bm r},t)=\sum\limits_{l=0}^{\infty}\sum\limits_{m=-l}^{l}\sum\limits_{n=1}^{\infty}
A_{lmn}j_l(k_{nl} r)Y_{lm}(\Omega)sin(\omega'_{nl}t)e^{-\tau_0t}.
\end{equation}

A solution of Eq. (\ref{e2100}) we represent as:
 \begin{equation}
  \delta \rho_2( {\bm r,t})=n_e f_2(r)+\sigma(\Omega,t)\delta(r+\epsilon -a), \;for\; r\geq a,\; ( r\rightarrow a+, i.e. \epsilon \rightarrow 0).
 \end{equation}

The neutrality condition, $\int\rho({\bm r},t)d^3r=N_e$, with
$\delta\rho_2({\bm r},t)=\sigma(\omega, t)\delta(a+\epsilon -r) +n_ef_2(r), (\;\epsilon\rightarrow 0)$, can be rewritten as follows:
$-\int\limits_{0}^{a} dr r^2f_1(r)= \int\limits_{a}^{\infty} dr r^2f_2(r)$, $\int\limits_{0}^{a}d^3rF({\bm r},t )=0$, $\int d\Omega \sigma(\Omega,t)=0$.
Taking into account also  the continuity condition on the spherical particle  surface, $1+f_1(a)=f_2(a)$, one can obtain:
$f_2(r)=\beta e^{-k_T(r-a)}/(k_Tr)$ and it is possible to fit $\alpha$ (cf. Eq. (\ref{alpha}))  and $\beta$ constants:
$\alpha=\frac{k_Ta +1}{2}$, $\beta=k_Ta -\frac{k_Ta+1}{2}\left(1-e^{-2k_Ta}\right)$---which gives Eqs (\ref{fff}).

 From the condition    $\int\limits_{0}^{a}d^3rF({\bm r},t )=0$  and  from Eq. (\ref{ea10}) it follows  that $A_{00n}=0$,
    (because of $\int d\Omega Y_{lm}(\omega)=4\pi \delta_{l0} \delta_{m0}$).

To remove the Dirac delta functions we integrate
both sides of the Eq. (\ref{e2100}) with respect to the radius ($\int\limits_{0}^{\infty}r^2dr...$)
 and then we take the limit to the sphere surface, $\epsilon\rightarrow 0$. It results in the following equation
 for surface plasmons:
\begin{equation}
\label{ea15}
\begin{array}{l}
\frac{\partial^2 \sigma(\Omega,t)}{\partial t^2}+\frac{2}{\tau_0}\frac{\partial \sigma(\Omega,t)}{\partial t}=- \sum\limits_{l=0}^{\infty}\sum\limits_{m=-l}^{l}  \omega_{0l}^2
Y_{lm}(\Omega)\int d\Omega_1  \sigma(\Omega_1,t)Y^{*}_{lm}(\Omega_1)\\
+\omega_p^2n_e  \sum\limits_{l=0}^{\infty}\sum\limits_{m=-l}^{l} \sum\limits_{n=1}^{\infty} A_{lmn}
\frac{l+1}{2l+1} Y_{lm}(\Omega)  \int\limits_{0}^{a}dr_1 \frac{r_1^{l+2}}{a^{l+2}}j_l(k_{nl}r_1)sin(\omega_{nl}t),\\
+\frac{en_e}{m}\sqrt{4\pi/3}\left[E_z({\bm R},t)Y_{10}(\Omega)+\sqrt{2} E_x({\bm R},t)Y_{11}(\Omega)   + \sqrt{2} E_y({\bm R},t)Y_{1-1}(\Omega)
\right],
\end{array}
\end{equation}
where $\omega_{0l}^2=\omega_p^2\frac{l}{2l+1}$.
 In derivation of the above equation the following formulae were exploited,
(for $a<r_1$):
    \begin{equation}
                    \frac{\partial}{\partial a}\frac{1}{\sqrt{a^2+r_1^2-2ar_1cos\gamma}}
                    =  \frac{\partial}{\partial a}  \sum\limits_{l=0}^{\infty}\frac{a^l}{r_1^{l+1}}P_l(cos\gamma)= \sum\limits_{l=0}^{\infty}
                    \frac{la^{l-1}}{r_1^{l+1}}P_l(cos\gamma),
                    \end{equation}
where $P_l(cos\gamma)$ is the Legendre polynomial [$P_l(cos\gamma)=\frac{4\pi}{2l+1}\sum\limits_{m=-l}^{l}
Y_{lm}(\Omega)Y^{*}_{lm}(\Omega_1)$], $\gamma$ is an angle between vectors ${\bm a}=a\hat{\bm r}$ and ${\bm r}_1$,
and (for $a>r_1$):
 \begin{equation}
                    \frac{\partial}{\partial a}\frac{1}{\sqrt{a^2+r_1^2-2ar_1cos\gamma}}
                    =  \frac{\partial}{\partial a}  \sum\limits_{l=0}^{\infty}\frac{r_1^l}{a^{l+1}}P_l(cos\gamma)=
                    -\sum\limits_{l=0}^{\infty}\sum\limits_{m=-l}^{l} 4\pi \frac{l+1}{2l+1} \frac{r_1^l}{a^{l+2}}
                    Y_{lm}(\Omega)Y^{*}_{lm}(\Omega_1).
                    \end{equation}

Taking into account the spherical symmetry, one can assume the solution of the Eq. (\ref{ea15}) in the form:
\begin{equation}
 \sigma(\Omega,t) = \sum\limits_{l=0}^{\infty}\sum\limits_{m=-l}^{l}q_{lm}(t)Y_{lm}(\Omega).
 \end{equation}
 From the condition $\int \sigma(\omega t)d\Omega=0$ it follows that $q_{00}=0$. Taking into account the initial condition
 $ \sigma(\omega, 0)=0$ we get (for $l\geq 1$),
 \begin{equation}
 \begin{array}{l}
 q_{lm}(t) =\frac{B_{lm}}{a^2}sin(\omega_{0l}'t)e^{-t/\tau_0}(1-\delta_{l1})+Q_{1m}(t)\delta_{l1}\\
 + \sum\limits_{n=1}^{\infty} A_{lmn}
\frac{(l+1)\omega_p^2}{l\omega_p^2-(2l+1)\omega_{nl}^2} n_e \int\limits_{0}^{a}dr_1
\frac{r_1^{l+2}}{a^{l+2}}j_l(k_{nl}r_1)sin(\omega_{nl}'t)e^{-t/\tau_0},\\
\end{array}
\end{equation}
where $\omega_{0l}'=\sqrt{\omega_{0l}^2-1/\tau_0^2}  $    and  $ Q_{1m}(t)$ satisfies the equation:
\begin{equation}
\frac{\partial^2 Q_{1m}(t}{\partial t^2}+\frac{2}{\tau_0}\frac{\partial Q_{1m}(t)}{\partial t} +\omega_{01}^2Q_{1m}(t)
= \frac{en_e}{m}\sqrt{4\pi/3}\left[E_z({\bm R},t)\delta_{m0}+\sqrt{2} E_x({\bm R},t)\delta_{m1}   + \sqrt{2} E_y({\bm R},t)\delta_{m-1}\right].
\end{equation}
Thus  $ \sigma(\omega, t)$ attains the form:
\begin{equation}
\begin{array}{l}
\sigma(\Omega,t)=\sum\limits_{l=2}^{\infty}\sum\limits_{m=-l}^{l} Y_{lm}(\Omega) \frac{B_{lm}}{a^2}sin(\omega_{0l}'t)e^{-t/\tau_0}+\sum\limits_{m=-1}^{1}
Q_{1m}(t)Y_{1m}(\Omega)\\
 +\sum\limits_{l=1}^{\infty}\sum\limits_{m=-l}^{l}
\sum\limits_{n=1}^{\infty} A_{nlm}
\frac{(l+1)\omega_p^2}{l\omega_p^2-(2l+1)\omega_{nl}^2}Y_{lm}(\Omega)
 n_e \int\limits_{0}^{a}dr_1 \frac{r_1^{l+2}}{a^{l+2}}j_l(k_{nl}r_1)sin(\omega_{nl}'t)e^{-t/\tau_0}.\\
\end{array}
\end{equation}

\end{document}